\documentclass[aps,twocolumn,showpacs,floatfix]{revtex4}
\usepackage{graphicx}
\usepackage{amsmath}
\usepackage{amsfonts}
\usepackage{amssymb}
\usepackage{epsfig}
\newcommand{\be}{\begin{equation}}
\newcommand{\ee}{\end{equation}}

\newcommand{\ket} [1] {\vert #1 \rangle}
\newcommand{\bra} [1] {\langle #1 \vert}
\newcommand{\im}{\textrm{i}}
\def\ie{{\it i.\ e.\ }}
\begin{document} 

\title{Simulation of the Laughlin state in an optical lattice}
\author{A. Riera}
\affiliation{Dept. d'Estructura i Constituents de la Mat\`eria,
Universitat de Barcelona, 647 Diagonal, 08028 Barcelona, Spain}

\date\today

\begin{abstract}
We analyze the proposal of achieving a Mott state of Laughlin wave functions in an optical lattice
[M. Popp {\it et al.}, Phys. Rev. A 70, 053612 (2004)] and
study the consequences of considering the anharmonic corrections to each single
site potential expansion that were not taken into account until now.
Our result is that, although the anharmonic correction reduces the maximum
frequency at which the system can rotate before the atoms escape from each site (centrifugal limit),
the Laughlin state can still be achieved for a small number of particles and 
a realistic value of the laser intensity.
\end{abstract}

\maketitle

\section{Introduction}
The fractional quantum Hall effect (FQHE) is one of the most 
studied phenomena in Condensed Matter Physics \cite{Tsui:1982-48}. 
Despite the fact that
a complete understanding of it is still missing, 
it is commonly believed that the interactions between 
the particles are essentially responsible for the 
strange states of matter that the 2D electron gas shows
at some particular values of the transverse magnetic field. 
In this respect, in 1983,
Laughlin proposed an Ansatz for the wave function of the ground state of the system \cite{Laughlin:1983-50}. 
This wave function, defined by
\begin{equation}
\label{eq:Laughlin_state}
\Psi_m(z_1,\ldots,z_N)\sim
\prod_{i<j}(z_i-z_j)^{m} 
e^{-\sum_{i=1}^{N}\vert z_i \vert^2/2} \ ,
\end{equation}
where $z_j=x_j+\textrm{i}y_j, j=1,\ldots,N$ 
stands for the position of the $j$-th particle, 
and $m$ is an integer number, describes the fractional quantum Hall 
state at a filling fraction $\nu=1/m$.
However, this state has only been proven to be an exact eigenstate 
of very specific Hamiltonians \cite{Yoshioka:2002-book} and for some specific 
values of the filling fraction; it contains the relevant properties 
that the ground state of the real system must have.

One of the most important features of some of the fractional quantum Hall states
is that they are states of matter whose quasiparticle 
excitations are neither bosons nor fermions, 
but particles known as {\it non-Abelian anyons}, 
meaning that they obey {\it non-Abelian braiding statistics}.
These new phases of matter define a new kind of order in Nature, 
a {\it topological} order \cite{Wen:2004-book}.
Such systems have become
very interesting from the
quantum computation perspective.
Quantum information could be stored in states 
with multiple quasiparticles, which would have a topological degeneracy. 
The unitary gate operations would be simply carried out by braiding quasiparticles, 
and then measuring the multi-quasiparticle states.
In this respect, let us also mention that several spin systems 
with topological order has been proposed recently
as candidates for robust quantum computing \cite{Kitaev:2006-321,Doucot:2005-71}. 
It has also been shown that these proposals
could be implemented by means of cold atoms \cite{Duan:2003-91,Micheli:2006-2}.

Despite the great possibilities of the FQH states,
neither the Laughlin wave function
nor its anyonic excitations have been observed directly in an experiment so far.
Nevertheless, recent experimental advances in the field of ultra-cold atomic gases
suggest that they could be good systems to simulate many Condensed Matter phenomena,
and, in particular, the FQHE. From the theoretical point of view, it has been shown that 
the FQH can be realized by 
rotating a bosonic cloud in a harmonic trap (see Ref.\ \cite{Paredes:2001-87, Paredes:2002-66}).
The rotation plays the role of the 
magnetic field for the neutral atoms. Thus, in the fast rotation
regime, the atoms live in the lowest Landau level and, 
if a repulsive interaction is introduced, they form the Laughlin wave function. 
In such system, the Laughlin state is a stable
ground state, however, in practice, due to the weak interaction between the particles,
the gap is too small, and it is not possible to achieve it experimentally.

An idea to avoid this problem is to use optical lattices \cite{Lewenstein:2006-56, Bloch:2008-80}. 
In such systems, the interaction energies are larger, since the atoms are
confined in a smaller volume. This yields a larger energy gap and, therefore, 
opens the possibility of achieving FQH states experimentally.

In the present work, we improve the existing proposal of 
achieving the Laughlin state in an optical lattice presented in Ref.\ \cite{Popp:2004-70}
by studying the consequences of considering not only
the harmonic approximation to the single site potential expansion,
but also the anharmonic corrections.
Our work is organized as follows: 
first, we analyze the one body Hamiltonian of our system. 
The solution of the 2D harmonic oscillator is reviewed and a more realistic model
for the single site potential of a triangular optical lattice is discussed. 
Then, we study the many particle problem by introducing a repulsive contact interaction
between the particles.
We compute, by means of the exact diagonalization technique, the fidelity of the ground state of the system
with the Laughlin wave function for a wide range of the experimental parameters. 
Finally, we discuss which are the required experimental conditions and the procedure to obtain
the Laughlin state.

\section{One body Hamiltonian}
\subsection{Harmonic case}
Let us consider one atom confined in a harmonic potential
which rotates in the $x-y$ plane at a frequency $\Omega$.
We will assume that the confinement in the $z$ direction is
sufficiently strong so that we can ignore the excitations
in that direction. The Hamiltonian associated with this system is
\begin{equation}
\hat{H}_0'=\frac{1}{2M}\left(\hat{p}_x^2 + \hat{p}_y^2\right)+
\frac{1}{2}M \omega^2(\hat{x}^2 +\hat{y}^2) - \Omega \hat{L}  ,
\end{equation}
where $M$ is the mass of the particle, $\hat{p}_x$ 
and $\hat{p}_y$ are the canonical 
momenta associated with the position coordinates $x$ and $y$, 
$\omega$ is the frequency of the trap and 
$\hat{L}$ is the angular momentum operator.

We can easily diagonalize this Hamiltonian by defining 
the creation and annihilation operators of 
some circular rotation modes
$\hat{a}_{\pm} =\frac{1}{\sqrt{2}}\left(\hat{a}_x \pm \im \hat{a}_y\right)$,
where $\hat{a}_{x,y}$
are the standard annihilation operators of the 1D harmonic oscillator in the directions $x,y$.
In terms of the number operators $\hat n_\pm\equiv \hat{a}_{\pm}^\dagger \hat{a}_{\pm}$
corresponding to these circular creation and annihilation operators, 
the angular momentum can be written as $\hat{L}=\hbar\left( \hat n_+ - \hat n_- \right)$
and the Hamiltonian reads
\begin{equation}
\hat{H}_0'= \hbar\omega (\omega-\Omega) \hat n_+
     + \hbar (\omega+\Omega) \hat n_-  +\hbar \omega \, .
\label{eq:original-Hamiltonian}
\end{equation}
Its spectrum is, therefore,
\begin{equation}
E^0_{n_+,n_-} = \hbar(\omega-\Omega)n_+ + \hbar(\omega+\Omega)n_- + \hbar\omega \, ,
\end{equation}
where $n_\pm$ are the integer eigenvalues of the number operators $\hat n_\pm$.

In the fast rotation regime ($\Omega \rightarrow \omega $), 
the family of states 
\begin{equation}
\ket{m}\equiv \ket{n_+=m, n_-=0}=\frac{1}{\sqrt{m!}}\left(a_+^\dagger \right)^m\ket{0} \, , 
\end{equation}
where $\ket{0}$ is the state annihilated by both $\hat a_+$ and $\hat a_-$,
form the subspace of lowest energy, usually known as the lowest Landau Level (LLL).
We can write the wave function of these states
\begin{equation}
\varphi_{m}(z) \equiv \langle z \ket{m}
=\frac{1}{\sqrt{\pi m!}}z^me^{-\frac{|z|^2}{2}} \, ,
\label{eq:monoparticle-basis}
\end{equation} 
where $\ket{z}=\ket{x,y}$ are the eigenstates of $\hat x$ and $\hat y$, 
$z=\frac{x+ \im y}{\ell}$ is a complex variable and 
$\ell= \sqrt{\frac{\hbar}{M \omega}}$ is a characteristic length of the system.

Notice that if the rotation frequency is high but smaller than
 the trap frequency, the lowest energy subspace is
only formed by states with $m < \frac{\omega + \Omega}{\omega-\Omega}$.
In this limit, we can find an effective Hamiltonian of our system by
projecting the original Hamiltonian onto the LLL, 
\begin{equation}
\hat{H_0}=\hbar(\omega-\Omega)\hat{L} \, .
\label{eq:effective-Hamiltonian}
\end{equation}
Indeed, although Hamiltonians of Eqs. (\ref{eq:original-Hamiltonian}) and 
(\ref{eq:effective-Hamiltonian}) are different, their projections onto the LLL
coincide, $P H_0' P^\dagger = P H_0 P^\dagger$, where $P=\sum_m \ket{m}\bra{m}$ is
the projector onto the LLL.

In Ref.\ \cite{Popp:2004-70}, the formation of fractional quantum Hall states in rotating optical
lattices is studied under the approximation of a single particle Hamiltonian as the presented
in Eq.\  (\ref{eq:effective-Hamiltonian}). 
Nevertheless, the potential of a site of an optical lattice is not infinite as the parabolic one.
Next, we model in a more realistic way this single site potential and
study how it will affect the formation of fractional quantum Hall states.

\subsection{Quartic correction}

\subsubsection{From the lattice potential to our model}
We will consider a system of bosonic atoms loaded 
in a 2D triangular optical lattice \cite{Becker:2009-ar}. 
The lattice is 2D due to the confinement in the $x-y$ plane created by two counter propagating
lasers in the $z$ direction. In each of this planes a triangular lattice is realized by
means of 3 lasers pointing at the directions
$\hat k_1=(\sin \theta,0, \cos \theta)$, $\hat k_2=(-\sin \theta /2, \sqrt{3} \sin \theta /2, \cos \theta)$
and $\hat k_3=(-\sin \theta /2, -\sqrt{3} \sin \theta /2, \cos \theta)$, where $\theta$ is the angle 
between these directions and the $z$ axis. 

In the region where the atoms are, the electric field created by 
each of these lasers can be modeled as a plane wave, 
$\vec E_j(\vec x, t)=A \vec \epsilon e^{\im (\vec k_j \cdot \vec x-\omega_L t)}$, 
where $\vec k_j=\frac{2\pi}{\lambda}\hat k_j$ for $j=1,2,3$ are 
the wave vectors of the lasers and $\lambda$ is the wavelength of the light used.
Notice that all the lasers share the same frequency $\omega_L$ and polarization vector $\vec \epsilon$.
The atoms are subject to the superposition of the electric field $\vec E_i(\vec x, t)$
created by each of the lasers, 
\be
\vec E(\vec x, t)=\sum_{i=1}^3\vec E_i(\vec x, t)= E(\vec x)\vec \epsilon e^{\im \omega_L t} .
\ee

After averaging over time, the effective potential that atoms feel is, therefore, the correction
to the energy of its internal state due to the AC-Stark effect.
This energy shift is proportional to the square of the amplitude of the electric field
$|E(\vec x)|^2$ and, in our particular case,
it can be written as
\begin{equation}
  U(\vec x)=- \frac{2}{9}U_0 \sum_{\langle i,j \rangle}
    \cos\left[(\vec k_i-\vec k_j)\vec x \right] \, ,
\label{eq:lattice-potential}
\end{equation}
where the sum runs over the 3 different couples of $ij=12, 23, 31$, 
and $U_0$ is the intensity of the laser. 
The minima of this potential form a Bravais triangular lattice with
basis vectors $\vec a_1= \frac{2}{3}\frac{\lambda}{\sin \theta}\left(-\frac{1}{2},\frac{\sqrt{3}}{2} \right)$
and $\vec a_2= \frac{2}{3}\frac{\lambda}{\sin \theta}(1,0)$.

Let us assume that the intensity of the laser $U_0 \gg E_R$, 
where $E_R=\hbar^2k^2/2M$ is the recoil energy, with $k=2\pi / \lambda$. 
In this limit, tunneling of atoms between different sites is forbidden, 
and the lattice can be treated as a system of independent wells. 
Thus, the ground state of the system is a product state of the state of each site
(Mott phase) and we can study the whole system by studying each site of the lattice
independently of the others.

In order to describe the potential of a single site,
the previous lattice potential in Eq.\  (\ref{eq:lattice-potential}) is expanded
around the equilibrium position $\vec x = \vec 0$,
\be
U(\vec x)\sim \frac{1}{2}M \omega^2(x^2 +y^2)
-\frac{3}{32}k^4 U_0 \left(x^2+y^2\right)^2 + O(x^6) \, .
\ee
We get a harmonic oscillator term with frequency $\omega = 2 \sqrt{U_0E_R}/ \hbar$ 
plus a quartic perturbation and higher order corrections. 
Notice that both terms of the expansion respect the rotational symmetry. 
This is actually the reason why we have considered a triangular lattice
instead of the simpler square one. In the square lattice, the circular symmetry
is broken in fourth order of the expansion, while in the triangular one, this
does not happen until the sixth order.
From now on, it will be considered that our single site potential is
adequately described by only these harmonic and quartic terms of the expansion of the potential.
This approximation is reasonable since it has already been assumed $U_0\gg E_R$.

Furthermore, by introducing some phase modulators 
into the lasers that form the lattice, 
it is possible to create time averaged potentials that 
generate an effective rotation of the single lattice sites. 
If this rotation $\Omega$ is close to $\omega$, 
the system is in the fast  rotation regime 
and we can obtain the low energy Hamiltonian proceeding as before. 
Written in units of $\hbar \omega$, it becomes
\begin{equation}
\hat H=\hbar \left(1-\frac{\Omega}{\omega}\right)\hat{L} -
\gamma \left(\frac{\hat x^2+\hat y^2}{\ell^2}\right)^2 \, ,
\label{eq:hamiltoninan-model}
\end{equation}
where $\gamma\equiv\frac{3}{32} \frac{U_0k^4 \ell^4}{\hbar \omega}=
\frac{3}{64}\sqrt{\frac{E_R}{U_0}}$ is a perturbation parameter. 
Notice that the perturbation parameter, 
$\gamma$, only depends on the intensity
of the laser in units of the energy recoil. 

In summary, we have described the whole sophisticated optical lattice by a set of independent
wells modeled by the simple Hamiltonian presented in Eq.\  (\ref{eq:hamiltoninan-model}). 
This effective Hamiltonian is
the same from Eq.\  (\ref{eq:effective-Hamiltonian}) 
plus a quartic correction term which will be responsible
for all the new physics that are going to be studied next. 

\subsubsection{Exact solution}
Next, we find the lowest energy eigenstates of the  
Hamiltonian presented in Eq.\  (\ref{eq:hamiltoninan-model}).
First of all, we realize that both $H$ and $H_0$ are rotationally invariant, 
therefore, $[H,H_0]=[H,L]=0$.
Thus, $H, H_0$ and $L$ must have a common eigenbasis, 
and this can only be $\{ \ket{m} \}$.
We determine the eigenvalues of $H$ by computing the expected values
\begin{equation}
E_m=\bra{m}H\ket{m}=\left(1-\frac{\Omega}{\omega} \right)m-\gamma (m+1)(m+2) \, ,
\label{eq:anharmonic-spectra}
\end{equation}
in units of $\hbar \omega$.

Let us point out that the commutator between the Hamiltonians
of the harmonic and anharmonic systems before being projected onto LLL,
$[H',H_0']$, is not zero.
Therefore, their lowest energy eigenstates only coincide in the fast rotation regime.

It is important, then, to establish which is the fast rotation regime for the Hamiltonian
(\ref{eq:hamiltoninan-model})
or, in other words, under which conditions our low energy description of the system
is correct.
To see this, we can compute, using perturbation theory, the first correction to the energy
of the second Landau level (NL) states  
\be
E_m^{NL}=(m+2)-\gamma(m+7)(m+2) \, .
\ee 
We realize that the LLL is a good description of the lowest energy states, \ie
 $E_m \ll E_m^{NL}$, if $3 \gamma (m+1) \ll 1$. 
In this regime,
the projection of the Hamiltonian onto the LLL is a good effective description of our system.
From now on, this condition will be always assumed.

\subsection{Maximum rotation frequency}
\label{sec:centrifugal-limit}
In the harmonic case, we realize that 
if the rotating frequency $\Omega$ exceeds 
the frequency of the trap $\omega$ (see Eq.\  (\ref{eq:effective-Hamiltonian})),
the lower energy states have an infinite angular momentum, 
\ie all the particles of the system are expelled from the trap.
This maximum frequency is usually known as the centrifugal limit.
What we are going to study now is, therefore, how the 
introduction of the quartic perturbation will affect
the centrifugal limit of the system.
In order to answer this question we realize two
different analysis that give quite similar results.

The first argument consists of a semi-classical interpretation of
the dynamics of the particle.
In the anharmonic system, we have a competition between
the attractive force of the harmonic trap and the repulsive
forces corresponding to the quartic correction and the fictitious centrifugal
force.
In the central region, the harmonic trap dominates and the particles are confined,
while if a particle was in the exterior region, it would be expelled.
The limit radius between these two regions corresponds to the maximum 
of the effective potential that includes both the trap and centrifugal terms. 
Furthermore, from the semi-classical point of view, each stationary state $\ket{m}$
follows a circular trajectory of radius $r\sim \sqrt{m}$ around the origin.
We expect then that those states whose associated radius is less than the
limit radius are bound states. 
Thus, we find that the maximum rotation frequency, or the centrifugal limit, depends
on the maximum angular momentum that we want to keep in the trap. 
This is a remarkable difference with respect to the harmonic case, where the centrifugal limit
was the same for any angular momentum state.
In particular and according to this semi-classical criterion, 
the centrifugal limit for a particle with an angular momentum equal or less
than $m_L$  is
\be
\Omega_L = \omega \sqrt{1-4\gamma m_L}\simeq \omega(1-2\gamma m_L) \, .
\ee

\begin{figure}
\scalebox{0.73}{\includegraphics{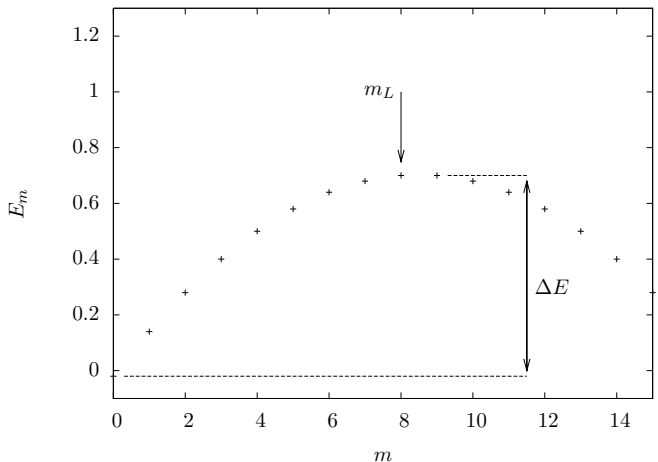}}
\caption{
Spectra of the single-particle Hamiltonian 
shown in Eq. (\ref{eq:hamiltoninan-model})
that models the dynamics of a particle in 
a rotating single site of a triangular optical lattice. 
This plot has been realized for $\gamma=0.01$ and $\Omega=0.8\omega$.
Notice that for these particular values of the frequency rotation and the laser intensity, 
particles with an angular momentum larger than $m_L=8$ would be expelled.
}
\label{figure:spectrum-sp}
\end{figure}

Another approach is to think in terms of the slope of the spectra $E_m$ 
presented in Eq.\  (\ref{eq:anharmonic-spectra}) and plotted in Fig.\ \ref{figure:spectrum-sp}.  
We realize that the single particle energy $E_m$ begins growing with $m$ from $m=0$, 
reaches a maximum at $\left[\frac{1-\Omega/\omega-3\gamma}{2\gamma}+\frac{1}{2}\right]$, 
(where the brackets $[$ $]$ represent the integer part function),
and becomes monotonically decreasing from this point.
We interpret this in the same way as the harmonic case, that is to say,
those states $\ket{m}$ that are in an energy interval with a negative slope 
are unstable.
According to this interpretation, the angular momentum of a particle must be
less than $\left[\frac{1-\Omega/\omega-3\gamma}{2\gamma}+\frac{1}{2}\right]$ in order to be trapped. 
This implies that
the centrifugal limit for a trapped particle with angular momentum $m_L$ is
\be
 \Omega_L =  \omega (1-2\gamma m_L - 4\gamma) \, .
 \label{eq:centrifugal-limit2}
\ee

Although the previous approaches are different, 
notice that they predict practically the same centrifugal limit.
In our simulations presented in Sec.\ \ref{sec:numerical-results}, we
have taken the condition given by Eq.\  (\ref{eq:centrifugal-limit2}), since
it is the most restrictive one.

At this point, we have completely solved the one particle problem.
Let us note that whereas in the harmonic case the maximum rotation frequency
coincides always with the trap frequency, 
in the system with the quartic perturbation,
this centrifugal limit depends on which is 
the maximum angular momentum of a particle that we want to keep trapped.
This difference will be crucial to understand why it will be more difficult
to obtain the Laughlin state in the anharmonic case than in the harmonic one.

\section{Many particle problem}
The Laughlin state is a strongly correlated state, therefore,
interaction between atoms will play an essential role in
its experimental realization.

Let us consider, then, a set of $N$ bosons trapped in a well
as the one described previously, and interacting
by means of a repulsive contact potential,
\begin{equation}
\label{contact_interaction}
V(\vec r - \vec r') = g \delta(\vec r - \vec r') \, ,
\end{equation}
where the parameter $g$ accounts for the strength of the interaction, 
and is related to the s-wave scattering length, $a_s$,
and to the localization length in the $z$ direction, $\ell_z$,
by $g=\sqrt{\frac{2}{\pi}}\frac{a_s}{\ell_z}$.

We are interested in knowing the ground state of the system
for a wide range of $g$ and $U_0/E_R$ in order to
see under which conditions the Laughlin state could be realized experimentally.
The solution of this problem is trivial in the extreme cases $g\to 0$ and $g\to\infty$.

When $g=0$, the atoms do not interact and the ground state of the system
is merely a product state with all the atoms with angular momentum zero.

If $g\to \infty$, the atoms will find a configuration in which they are
always in a different position with the minimum total angular momentum possible.
This configuration that minimizes the interaction energy at the expense of 
the angular momentum of the particles is precisely the Laughlin state.
As we can see in Eq.\  (\ref{eq:Laughlin_state}), 
the interaction energy in the Laughlin wave function is strictly zero 
since the probability that two atoms are in the same position is null.

For a finite $g$, the problem of finding the ground state of the system
has to be addressed numerically. In particular, we will solve it by means of
 exact diagonalization. 
With this aim, we can write the Hamiltonian of this system 
in second quantization form,
\begin{align}
\hat H &=\sum_{j=1}^L \left((1-3\gamma-\Omega)\hat{n}_i-\gamma \hat{n}_i^2+1-2\gamma\right)\nonumber \\ 
 &+\frac{1}{2}\sum_{i<j}V_{ijkl} b_ib_j b_k^\dagger b_l^\dagger \, ,
\label{eq:many-body-hamiltonian}
\end{align}
where $b_j$ and $b_j^\dagger$ are the creation and 
annihilation operators of the mode $j$ which corresponds to the LLL 
state with angular momentum $j$ presented in Eq.\  (\ref{eq:monoparticle-basis}), 
$\hat{n}_j\equiv b_j^\dagger b_j$ counts the number of particles with angular
momentum $j$, and $V_{ijkl}$ are the coefficients of the interaction in this basis
defined by
\begin{align}
V_{m_1\,m_2\,m_3\,m_4}&=\bra{m_1\,m_2} \hat V \ket{m_3\,m_4} \\
&=\frac{g}{2\pi}\,\,\frac{\delta_{m_1+m_2,
m_3+m_4}}{\sqrt{m_1!m_2!m_3!m_4!}}\,\,
\frac{(m_1+m_2)!}{2^{m_1+m_2}}\: . \nonumber
\end{align}
The cylindrical symmetry of the Hamiltonian allows the
diagonalization to be performed 
in different subspaces of well defined total $z$-component of
$L=\sum_{i=1}^N m_i$.

Thus, given the parameters $\Omega, U_0/E_R$ and $g$
and a subspace of total angular momentum $L$, 
we construct the multi-particle basis of $N$ particles compatible with $L$, 
calculate the matrix-elements of the Hamiltonian in this subspace
by means of Eq.\  (\ref{eq:many-body-hamiltonian}), 
and perform its diagonalization. 
Once the ground state for each subspace $L$
is determined, we find the ground state of the system
by selecting the state with lowest energy.

An important issue that we have to take into account
when we perform the simulation is to be careful with
the centrifugal limit.
As it has been shown in Sec.\ \ref{sec:centrifugal-limit}, 
if we want that the particle with the maximum angular momentum 
does not escape from the trap, we have to keep the rotation below
the centrifugal limit given by Eq.\  (\ref{eq:centrifugal-limit2}).
In our case, in which we want to drive the system into the Laughlin state,
this centrifugal limit is determined by the maximum angular momentum of a single
particle in the Laughlin state, that is $N(N-1)/2$. 
Thus, the maximum rotation frequency is given by Eq.\  (\ref{eq:centrifugal-limit2})
taking 
\be
m_L=N(N-1)/2 \, .
\ee

\section{Results and discussion} 
\label{sec:numerical-results}
In what follows, we display the results obtained from exact diagonalization
described in the previous section.

\begin{figure}
\scalebox{0.72}{\includegraphics{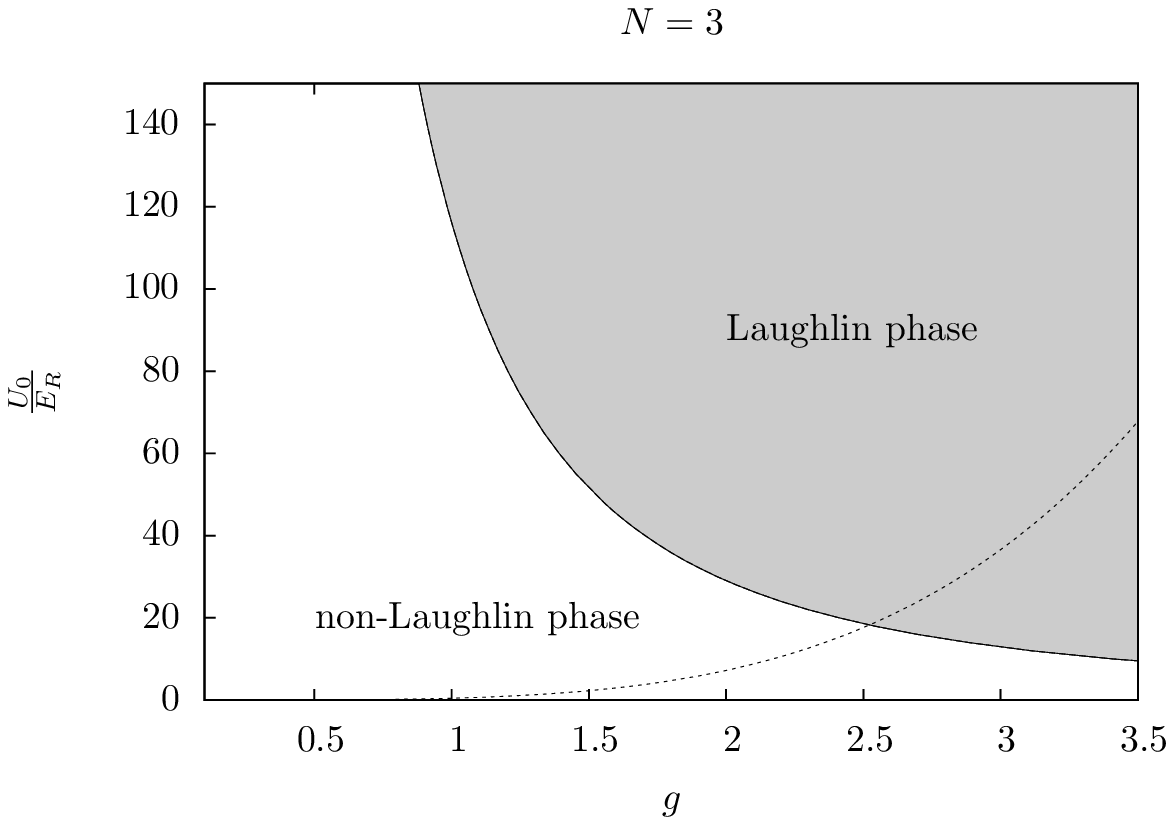}}\\
\vspace{1cm}
\scalebox{0.72}{\includegraphics{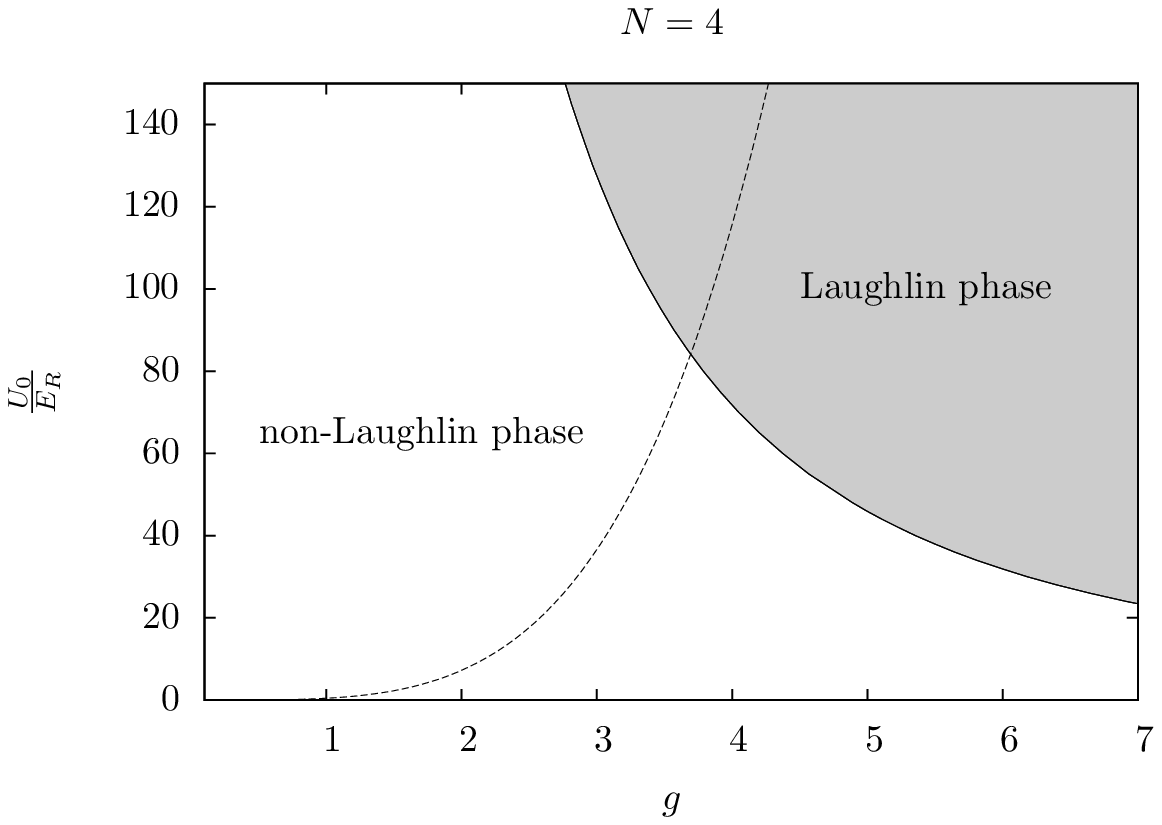}}\\
\vspace{1cm}
\scalebox{0.72}{\includegraphics{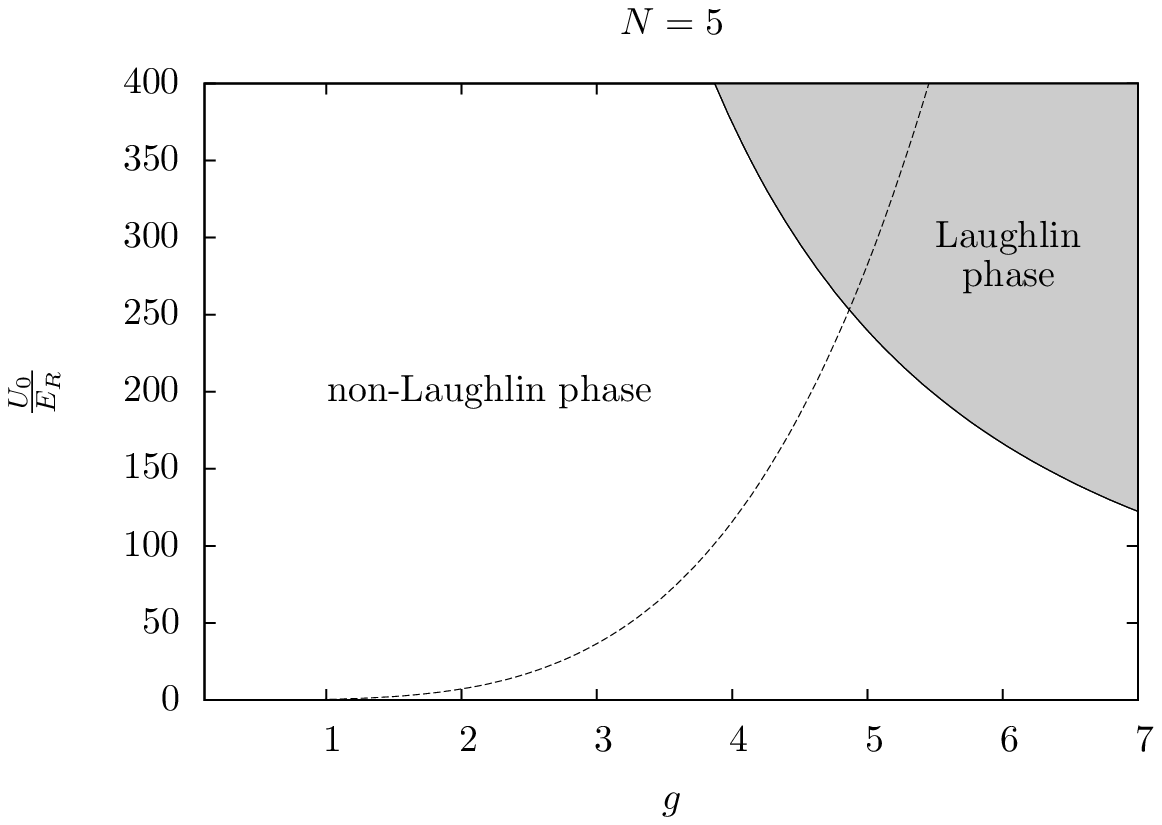}}\\
\caption{
Density plot of the fidelity between the GS of the system and 
the Laughlin wave function versus the laser intensity $U_0$
and the strength of the contact interaction $g$ for different
number of particles ($N$=3,4 and 5). In order to realize the diagram,
the maximum possible rotation frequency $\Omega_L$ has been considered
for each value of $U_0$.
The dashed line represents the dependence of $g$ on the confinement
$U_0$ for the Rubidium atoms case.
Although the $U_0$ needed to get the Laughlin as 
the ground state of the system
increases enormously with the number of particles, 
we realize that for small values of $N$, the laser
intensity required is achievable. 
}
\label{figure:density-plot}
\end{figure}

In order to study under which conditions the Laughlin state is 
the ground state of the system
a phase diagram is presented in Fig.\ \ref{figure:density-plot}.
The fidelity between the Laughlin
state and the GS of the system is plotted versus
the laser intensity $U_0/E_R$ and the strength of the
contact interaction $g$. 
For each value of $U_0/E_R$, we have taken 
the maximum possible rotation frequency, 
defined by Eq.\  (\ref{eq:centrifugal-limit2}), since it 
corresponds to the best condition to achieve the Laughlin wave function.
The plot shows two separated phases. In one of them, 
the Laughlin state is perfectly obtained 
($\bra{GS}\Psi_L \rangle=1$). It corresponds to
high values of the laser intensity and strength of the interaction. 
On the contrary, for low values of $U_0/E_R$ and $g$ the fidelity between
the GS and the Laughlin state is zero.
This phase corresponds to other 
states with less angular momenta
and not so correlated.

\begin{figure}
\scalebox{0.72}{\includegraphics{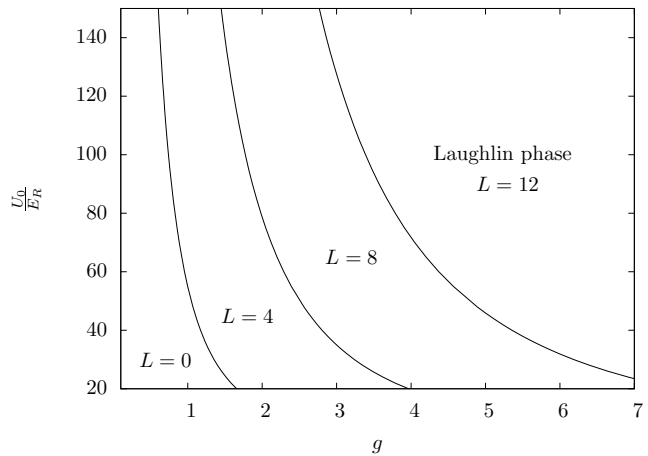}}\\
\caption{
Density plot of the angular momentum of the GS of the system respect
 to the laser intensity $U_0$
and the strength of the contact interaction $g$ for $N$=4. 
In order to realize the diagram,
the maximum possible rotation frequency $\Omega_L$ has been considered
for each value of $U_0$.
}
\label{figure:angular-momentum}
\end{figure}

Notice that the transition between 
the Laughlin and non-Laughlin phases is very abrupt.
This is because the ground state in the non-Laughlin phase has
an angular momentum smaller than the angular momentum that 
Laughlin state requires. 
States with different angular momentum are orthogonal and therefore
the fidelity between them is zero.
Nevertheless, when interaction is high enough and
it is sufficiently favorable for the system to have total angular momentum
$L=N(N-1)$, the Laughlin wave function is instantly achieved.
This argument is illustrated in Fig.\ \ref{figure:angular-momentum} for the case
of $N=4$ particles.
We can see how the system increases its
angular momentum depending on the confinement $U_0$ and the interaction $g$.

In Fig.\ \ref{figure:density-plot},  the shape of the boundary
between the Laughlin and non-Laughlin phases can be easily explained.
According to the argument presented in the previous section, 
the configuration of the ground state of the system depends 
on the ratio between the strength of the interaction $g$ and
the energy difference of the single particle spectrum
$\Delta E \equiv E_{m_L}-E_0=\gamma m_L(m_L+1)$ (see Fig.\ \ref{figure:spectrum-sp}).
In particular, we expect two regimes:
\begin{itemize}
\item If $g \gg \Delta E$, the GS is the Laughlin state,
\item while if $g \ll \Delta E$, then GS is a product state 
with all the particles with angular momentum 0.
\end{itemize}
The transition, then, takes place between these two regimes, that is, when $g \sim \Delta E$.
This condition implies that
\be
g = f(N) \sqrt{\frac{E_R}{U_0}}
\label{eq:fit}
\ee
where $f(N)\sim m_L(m_L+1)$ is a function that depends on $N$.

\begin{figure}
\scalebox{0.72}{\includegraphics{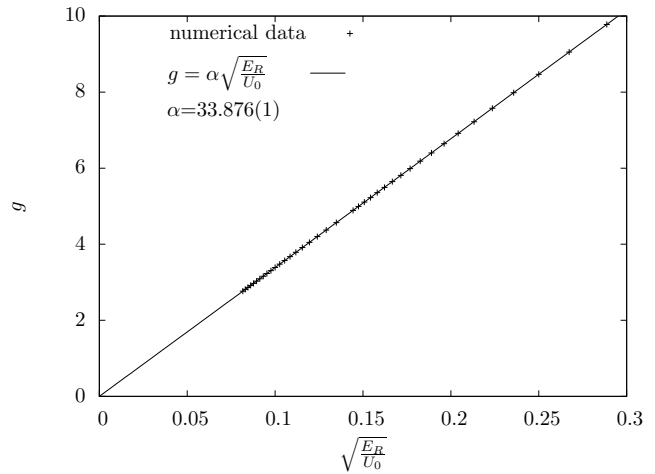}}\\
\caption{
Plot of the points $(\sqrt{\frac{E_R}{U_0}}, g)$ of the border between the two phases 
Laughlin and non-Laughlin of Fig. \ref{figure:density-plot} for $N=4$.
We realize that Eq. \ref{eq:fit} describes perfectly its behavior.
}
\label{figure:fit}
\end{figure}

In Fig.\ \ref{figure:fit}, we plot the numerical data $(\sqrt{\frac{E_R}{U_0}}, g)$ 
corresponding to the points of the
border between the Laughlin and non-Laughlin phases of Fig.\ \ref{figure:density-plot}
for $N=4$ in order to see if Eq.\ (\ref{eq:fit}) is fulfilled. 
We observe a perfect fit between numerical data and Eq.\ (\ref{eq:fit}) and,
performing a linear regression, we determine the slope $f(4)=33.876(1)$.
The same perfect agreement has been found for $N=3$ and $N=5$ cases, 
with slopes $f(3)=10.779(1)$ and $f(5)=77.427(3)$ respectively.

According to Eq.\ (\ref{eq:fit}), we would expect that function 
$f(N)$ scales as $\sim N(N-1)\left(N(N-1)+2\right)$.
Nevertheless, although $f(3)$, $f(4)$ and $f(5)$ seem to scale in this way, 
three points are not enough to confirm this behavior.


The dashed line in Fig.\ \ref{figure:density-plot} expresses the dependence of $g$ on the confinement
$U_0$ for the Rubidium atoms case (see Eq.\  (\ref{eq:g-vs-V0})). 
It shows the natural path that the system would follow increasing the
intensity of the laser without improving the strength of the interaction
by means of Feshbach resonance techniques.
This illustrates the advantages
of confining ultra-cold gases in an optical lattice.
The high confinement increases the scattering length
of the interaction between the atoms, 
\begin{equation}
g=\sqrt{2}a_s |k| \left(\frac{U_0}{E_R}\right)^{1/4} \, ,
\label{eq:g-vs-V0}
\end{equation}
which, at the end of the day, is responsible 
for the high correlation of the GS.

Moreover, this enhanced interaction produces
a larger energy gap which makes the optical lattice proposal more
robust compared to the harmonic trap setup. This enhancement in the energy
gap of the excitation spectrum can alleviate some of the challenges for
experimental realization of the quantum Hall state for ultra-cold
atoms.

In Fig.\ \ref{figure:density-plot}, 
we also realize that the larger the number of atoms of the system is,
higher the laser intensity needed to achieve the Laughlin state.
The reason for this is that for larger $N$, the maximum angular momenta
allowed before the atoms escape from the trap is larger, $m_L=N(N-1)/2$.
This implies a smaller maximum rotation frequency $\Omega_L$ (see Eq.\ (\ref{eq:centrifugal-limit2})),
and the requirement of increasing
the laser intensity $U_0$ to compensate this effect. 
Let us note that increasing $U_0$ we achieve both 
larger maximum rotation frequency and a stronger interaction $g$.

It is interesting to point out that driving 
the system into the Laughlin wave function is not possible
by simply increasing the rotation frequency, 
since the symmetric shape of the potential conserves the angular
momenta and the system would rest in a
trivial non-entangled state with angular momentum zero. 
To avoid this, we should introduce some deformation to the
wells of the lattice in order to break the spherical symmetry.
For our optical lattice
setup this can  be achieved by introducing
a couple of electro-optical fast phase modulators
whose effect in the rotating frame would be
a new trapping potential with the form $V_p \propto \left(\omega
+\Delta \omega\right)^2 x^2  + \omega^2 y^2$. 
This modification would change our original Hamiltonian in 
a new one $H+H_\epsilon$, where 
\begin{equation}
\label{eq:Hasymmetry}
H_{\epsilon}=\frac{\epsilon}{4} \sum_m \beta_m b_{m+2}^\dagger
b_{m}^{} + (m+1)  b_{m}^\dagger   b_{m}^{} + \textrm{h.c.}, 
\end{equation}
where $\beta_m=\sqrt{(m+2)(m+1)}$ and $\epsilon=\Delta \omega/
\omega$ is a small parameter.

The previous perturbation (\ref{eq:Hasymmetry}) leads
to quadrupole excitations, so that states whose total angular
momentum differ by two units are coupled, 
and the system can increase its angular momentum.
In order to drive the system into the Laughlin state, 
we could follow some adiabatic paths in
the parameter space ($\Omega$, $\epsilon$) such that the gap along this paths is
as large as possible, and in this way, 
keep the system in its ground state
according to the adiabatic theorem \cite{Born:1928}.

Finally, let us briefly discuss how to check that we have 
driven the system to the Laughlin state.
First, notice that as the system is in a Mott phase,
any measurement signal will be enhanced by a factor equal to
the number of occupied lattice sites. 

According to Ref.\ \cite{Read:2003-68}, for any LLL state of a trapped, rotating, interacting Bose gas, 
the particle distribution in coordinate space in a time of flight experiment is related to that 
in the trap at the time it is turned off by a simple rescaling and rotation.
Thus, it is possible to measure the density profile of our state by accomplishing a time of flight experiment.
Notice that by means the density profile it is possible to estimate the total angular momentum of the system.

Although the angular momentum of the system can be measured, 
there are many states with the same angular momentum
that are not the Laughlin state. Thus, in order to really distinguish if the system is in the Laughlin state,
the measurement of other properties is required. 
One possibility is the measurement of correlations. 
In Ref.\ \cite{Popp:2004-70}, an interesting technique is proposed to measure
the correlation functions $g_1=\langle \psi^\dagger ({\bf r})  \psi ({\bf r'})\rangle$ and
$g_2=\langle \psi^\dagger ({\bf r})  \psi^\dagger ({\bf r'})
\psi({\bf r})\psi ({\bf r'}) \rangle$.
These functions are very particular for the Laughlin state.
For instance, $g_2\propto \lvert
r-r'\vert^4$, since in the Laughlin state all the particles
have a relative angular momentum $m_r=2$ between them.
Let us just mention the scheme of this technique.
It lies in using a gas of two species that can be coupled via Raman transitions (hyperfine levels).
First, all the atoms are in the same state and the system is driven to the Laughlin state.
Then, an equal superposition of the two species is created by means of a
$\pi/2$-pulse with the laser. 
Next, the atoms are shifted a small distance compared to the lattice spacing
by moving the lattice of one of the species (as was proposed 
in Ref.\ \cite{Jaksch:2000-85} and experimentally realized in Ref.\ \cite{Mandel:2003-91}).
Finally, another $\pi/2$-pulse is performed in order to put all the atoms in the same state
and the time of flight measurement is accomplished.

\section{Conclusions}
The method for achieving the Laughlin state in an optical lattice
presented in Ref.\ \cite{Popp:2004-70} has been studied.
It has been shown that it is essential to take into account the anharmonic corrections.
A quartic anharmonic term introduces a maximum rotation frequency that is smaller than in the harmonic
case and, therefore, if this was not considered all the particles would be expelled from their 
single lattice sites.

Although this more restrictive centrifugal limit makes more difficult
to drive the system into the Laughlin state, since it requires higher values of $U_0$, for
systems with a small number of particles, the Laughlin state is achievable experimentally.

The shape of the boundary between the Laughlin and non-Laughlin phases in the
phase diagram of the system respect to the confinement $U_0$ and the strength of the interaction $g$
plotted in Fig.\ \ref{figure:density-plot} has been physically explained and analytically described.

\begin{acknowledgments}
The main part of this work has been realized under the supervision of B. Paredes.
Let us also thank S. Trotzky, S. Iblisdir and M. Moreno for fruitful
discussions. 
Financial support from QAP (EU), MICINN (Spain), FI
program and Grup Consolidat (Generalitat de Catalunya), and
QOIT Consolider-Ingenio 2010 is acknowledged.
\end{acknowledgments}

\bibliography{Laughlin}

\end{document}